\pageno=1 
\magnification=\magstep1
\def\bs{\bigskip}
\def\ni{\noindent} 
 
\def\eq{\eqno}
\input epsf

\bs\bs 

\centerline {\bf NEUTRON STARS IN RELATIVISTIC MEAN FIELD THEORY} 

\centerline {\bf WITH ISOVECTOR SCALAR MESON} 

\bs\bs \bs\bs

\centerline {S.  Kubis$^a$, M.  Kutschera$^{a,b}$ and S. Stachniewicz$^a$} 

\centerline {$^a$~H.Niewodnicza\'nski Institute of Nuclear Physics} 

\centerline {ul. Radzikowskiego 152, 31-342 Krak\'ow, Poland} 

\centerline {$^b$~Institute of Physics, Jagellonian University}

\centerline {ul. Reymonta 4, 30-059 Krak\'ow, Poland}

\bs \bs \bs \bs 

\ni Abstract:

We study
the equation of state (EOS) of $\beta$-stable dense matter and
models of neutron stars in the
relativistic mean field (RMF) theory  with the isovector
scalar mean field corresponding to the $\delta$-meson
[$a_0(980)$]. A range of values of the $\delta$-meson coupling
compatible with the Bonn potentials is explored. 
Parameters of the model in the isovector sector are constrained
to fit the nuclear 
symmetry energy, $E_s \approx 30 MeV$. We find that the quantity
most sensitive to the $\delta$-meson coupling is the proton
fraction of neutron star matter. It increases significantly in
the presence 
of the $\delta$-field. The energy per baryon also increases but
the effect is smaller.
The EOS becomes slightly stiffer  and
the
maximum neutron star mass increases  for stronger
$\delta$-meson coupling.

\ni PACS: 21.65.+f, 97.60.Jd


\bs \bs \bs

\ni {\bf 1. Nucleon matter in the RMF model with the $\delta$-meson}

 The standard RMF model [1] of nuclear matter, frequently used in
astrophysical calculations, involves mean fields of  $\sigma$,
$\omega$ and $\rho$ mesons. It does not include the contribution
of the isovector scalar meson $\delta$ [$a_0(980)$]
although generally the density to which
this field can couple does not vanish, $<{\bar\psi}\tau_3\psi> \ne 0$.
The contribution of the $\delta$-meson field is not expected to be
important for finite
nuclei of small isospin-asymmetry, as the $\delta$-meson
mean field vanishes in symmetric 
nuclear matter. However, for strongly isospin-asymmetric matter
in neutron stars presence of the $\delta$-field can influence
the properties of dense matter.
In Ref.[2]  the RMF model was generalized to
include the contribution of the $\delta$-meson. Here we 
investigate consequences of such a generalized RMF theory for
neutron stars.

The dynamics of the RMF model is governed by the lagrangian

$$ L=L_0+L_{int}, \eq(1)$$

\ni where $L_0=L_{\psi}+L_{\sigma}+L_{\omega}+L_{\rho}+L_{\delta}$
is the free-field lagrangian and $L_{int}$ is 
the interaction term.
The free-field lagrangians for nucleons and meson fields are:

$$ L_{\psi}= {\bar \psi}(i \partial_{\mu} \gamma^{\mu}-m)\psi,
\eq(2) $$
\ni where $m$ is the bare nucleon mass,

$$ L_{\sigma}={1 \over 2}(\partial_{\mu} \sigma\partial^{\mu}
\sigma - m_{\sigma}^2 \sigma^2), \eq(3)$$

$$ L_{\omega}=-{1 \over 4}(\partial_{\mu}\omega_{\nu} - \partial_{\nu}\omega_{\mu})
(\partial^{\mu}\omega^{\nu} - \partial^{\nu}\omega^{\mu})
+{1 \over 2}m_{\omega}^2 \omega_{\mu}\omega^{\mu}, \eq(4)$$

$$ L_{\rho}=-{1 \over 4}(\partial_{\mu}{\vec \rho}_{\nu} -\partial_{\nu}{\vec
\rho}_{\mu})(\partial^{\mu}{\vec \rho}^{\nu} -\partial^{\nu}{\vec
\rho}^{\mu})
+{1 \over 2}m_{\rho}^2{\vec \rho}_{\mu}{\vec
\rho}^{\mu}. \eq(5)$$  

\ni Here $m_{\sigma}$, $m_{\omega}$, and $m_{\rho}$ are masses of
respective mesons. Lagrangians (2)-(5) are the same as used in
the standard RMF theory [1].

 For the $\delta$-field
we use the simplest lagrangian of the massive isovector scalar field,

$$L_{\delta}={1 \over 2} \partial_{\mu}{\vec
\delta}\partial^{\mu}{\vec \delta}-{1 \over 2}m_{\delta}^2{\vec
\delta}^2,  \eq(6)$$

\ni where $m_{\delta}$ is the mass of the
$\delta$-meson.

The coupling of the meson fields to nucleons is assumed to have
the Yukawa form.
 For the
$\sigma$-field we use the cubic and quartic selfinteraction
terms.
The interaction lagrangian reads

$$ L_{int}=g_{\sigma} \sigma {\bar \psi} \psi - g_{\omega}
\omega_{\mu} {\bar \psi} \gamma^{\mu} \psi -{1 \over 2}g_{\rho} {\vec \rho}_{\mu}
{\bar \psi} \gamma^{\mu} {\vec \tau} \psi +g_{\delta} {\vec \delta}
{\bar \psi}{\vec \tau} \psi - U(\sigma),  \eq(7)$$

\ni where $U(\sigma)$ is the potential energy term of the
$\sigma$-field due to Boguta
and Bodmer [3],

$$ U(\sigma)= {1 \over 3} b m\sigma^3 + {1 \over 4} c \sigma^4.
\eq(8)$$

In the RMF approximation, for a uniform nucleon matter only a few
fermion densities are relevant. These include
the baryon density, $<{\bar \psi} \gamma_0 \psi>=n_B$, 
the scalar density, 

\ni $<{\bar\psi}\psi>$, the isospin density,
$<{\bar\psi}\gamma_0 \tau_3 \psi>$, and the scalar 
isospin density, 
$<{\bar\psi}\tau_3\psi>$.
A selfconsistent description of the system is achieved by
taking into account only 
those components of the meson fields which couple to the above
densities with all remaining components vanishing.
For normal nucleon
matter  the relevant components of mean meson fields are ${\bar
\sigma}$, ${\bar \omega}_0$,  
${\bar \rho}_0^{(3)}$, and,
for the $\delta$-field, the isospin component ${\bar \delta}^{(3)}$.

It is a simple algebraic exercise to obtain the spectrum of
nucleon energies in terms of the above components of meson
fields and to construct all relevant nucleon densities. 
The single particle energies of protons and neutrons are 

$$ E_{P(N)}({\bf p})=g_{\omega}{\bar \omega}_0 \pm {1 \over 2} 
g_{\rho}{\bar \rho}^{(3)}_0+ \sqrt{{\bf p}^2+m_{P(N)}^2}~~, \eq(9)$$

\ni where the proton and neutron effective mass is, respectively,

$$ m_P=m-g_{\sigma}{\bar \sigma}-g_{\delta}{\bar \delta^{(3)}}, \eq(10) $$

\ni and

$$ m_N=m-g_{\sigma}{\bar \sigma}+g_{\delta}{\bar \delta^{(3)}}. \eq(11) $$

\ni The plus (minus) sign in the formula (9) refers to protons (neutrons).

We  obtain
the field equations for mean meson fields in the form:

$$ m_{\omega}^2{\bar \omega}_0=g_{\omega}n_B, \eq(12)$$

$$ m_{\rho}^2{\bar \rho}_0^{(3)}={1 \over 2} g_{\rho} (2x-1)n_B, \eq(13)$$

\ni  where 
$x=n_P/n_B$ is the proton fraction,

$$m_{\sigma}^2{\bar \sigma}+{\partial U \over \partial \sigma}=g_{\sigma}(n^s_P+n^s_N), \eq(14)$$

\ni and

$$m_{\delta}^2{\bar \delta^{(3)}}=g_{\delta}(n^s_P-n^s_N).  \eq(15)$$

\ni In Eqs.(14) and (15) $n^s_P$ and $n^s_N$ is, respectively, proton and neutron
scalar density,

$$ n^s_i={2 \over (2\pi)^3} \int^{k_i}_0d^3k {m_i \over
\sqrt{k^2+m_i^2}}, ~~~~~~~~i=P,N,\eq(16)$$

\ni with proton and neutron effective mass, $m_P$ and
$m_N$, given in Eqs.(10) and (11). Mean meson fields as
determined through Eqs.(12)-(15) depend on the baryon density $n_B$ and 
the proton fraction $x$.

The energy of the uniform matter consists of nucleon and meson
contributions. The nucleon contribution is a sum of proton and
neutron energies (9) up to their respective Fermi momenta. Mean meson
field contributions are easily obtained from the lagrangians (4)-(6).
With the mean meson fields given by Eqs.(12)-(15)
the energy density of a uniform nucleon matter becomes

$$ \epsilon_{nuc} = {2 \over (2\pi)^3} (\int^{k_P}d^3k
\sqrt{k^2+m_P^2}+\int^{k_N}d^3k \sqrt{k^2+m_N^2})+ {1 \over
2}C_{\omega}^2n_B^2 + $$

$$ +{1 \over 2} {1 \over C_{\sigma}^2}[m-{m_P+m_N \over 2}]^2+ U({\bar
\sigma})+{1 \over 8}C_{\rho}^2(2x-1)^2n_B^2+{1 \over 8} {1 \over
C_{\delta}^2}(m_N-m_P)^2.  \eq(17)$$

In the spirit of the RMF theory [1] the parameters of the model
are fit to reproduce the empirical parameters of nuclear
matter.
In the isoscalar sector, the coupling parameters,
$C_{\sigma}^2 \equiv g_{\sigma}^2/m_{\sigma}^2,
C_{\omega}^2 \equiv g_{\omega}^2/m_{\omega}^2$, ${\bar b} \equiv
b/g_{\sigma}^3$, and ${\bar c} \equiv c/g_{\sigma}^4$, are adjusted to
fit the saturation properties of symmetric nuclear matter ($x=1/2$), i.e.
the saturation density $n_0=0.145fm^{-3}$, the binding energy $B=-16
MeV$ per nucleon, and the compressibility modulus $K_V \approx 280MeV$.
The fourth parameter, e.g.  ${\bar c}$, can be used to control the stiffness of
the equation of state of symmetric nuclear matter.  Our choice
of these  coupling parameters is discussed below. 

The coupling parameters, $C_{\rho}^2 \equiv g_{\rho}^2/m_{\rho}^2$, and
$C_{\delta}^2 \equiv g_{\delta}^2/m_{\delta}^2$, in the isovector 
sector are constrained to fit the 
nuclear symmetry energy, $E_s=31\pm 4 MeV$ [4].  
This constraint gives $C_{\rho}^2$ as a function of $C_{\delta}^2$ [2].
In the range of values of $C_{\delta}^2$ 
considered below, which is compatible with the Bonn potentials
[5], this function is well approximated by a linear relation [2]

$$C_{\rho}^2=AC_{\delta}^2+B, \eq(18)$$

\ni with positive coefficients, $A>0$, and $B>0$, depending weakly on the coupling
parameters in the isoscalar sector.

The requirement that the nuclear symmetry energy is reproduced
in the presence of the $\delta$-field is very important as the $\delta$-field
contribution is strongly attractive. As shown in Ref.[2] the
$\delta$-meson coupling provides a negative contribution to the
symmetry energy that tends to cancel partially the $\rho$-meson
contribution. Thus to preserve the nuclear symmetry energy a
stronger $\rho$-meson coupling is needed.
 The formula (18) shows that the parameter
$C_{\rho}^2$ has the lowest value
$C_{\rho}^2=B$ for $C_{\delta}^2=0$. 
For any coupling constant $C_{\delta}^2> 0$, the
value of the $\rho$-coupling constant, $C_{\rho}^2$, increases.
Hence for pure 
neutron matter inclusion of
the $\delta$-meson results 
unavoidably in higher energy per particle at high
densities, where contributions of vector mesons dominate.

\bs
\ni{\bf 2. Equation of state and neutron stars}

As we mentioned above, there is one parameter, e.g. ${\bar c}$, which
can be used to label a family of EOS's in the RMF model. We wish
to retain this freedom, as the true high density behaviour of
the neutron star EOS is only weakly constrained at present [6]. 
In the following
we shall mainly use two sets of coupling parameters which
reproduce the saturation  
properties of nuclear matter but predict different stiffness of the EOS at higher
densities.  The EOS's corresponding to these two sets of
parameters 
are referred to as soft and stiff. 
 The soft
EOS is specified by the parameters
$C_{\sigma}^2=1.582fm^2, C_{\omega}^2=1.019fm^2, {\bar
b}=-0.7188$, and ${\bar c}=6.563$.
For the stiff EOS  the parameters are
$C_{\sigma}^2=11.25fm^2, C_{\omega}^2=6.483fm^2, {\bar b}=0.003825$, and
${\bar c}=3.5\times10^{-6}$. 

The above EOS's limit in a sense the family of EOS's in the
RMF model. They are close to the
softest and to the stiffest  EOS  in the RMF
theory that are physically allowed. The stiff EOS corresponds to
a very small value of the parameter ${\bar c}$. Physical
consistency of the RMF theory requires that ${\bar c}>0$.  In
Fig.1 we show a plot 
of the parameter ${\bar c}$ as a 
function of $C_{\sigma}^2$.
As one can see,
${\bar c}=0$ for $C_{\sigma}^2 \approx 11.5 fm^2$ and thus
any acceptable value of $C_{\sigma}^2$ must be lower (Fig.1). In
terms of the maximum neutron star mass, the EOS with ${\bar c}
\to 0$ is
the stiffest acceptable EOS in the RMF model. 
We choose small but finite value
${\bar c}=3.5\times10^{-6}$ with corresponding
$C_{\sigma}^2=11.25fm^2$. 

The parameter ${\bar c}$ for the soft EOS has the highest allowed value, 
${\bar c}=6.563$.    
The maximum
neutron star mass corresponding to this EOS is about $1.44
M_{\odot}$. This is the mass of the 
heavier neutron star in the binary pulsar PSR B1913+16 [7], which has the
largest precisely measured value of a neutron star mass. Hence this EOS is the 
softest one still compatible with measured neutron star masses.

We also show below some results for an intermediate EOS with 
the parameters
$C_{\sigma}^2=5.318fm^2, C_{\omega}^2=2.31fm^2, {\bar b}=-0.03952$, and
${\bar c}=0.4229$. 

The coupling constant $g_{\delta}$ of the $\delta$-nucleon
interaction is a parameter in the one-boson-exchange fits of nucleon-nucleon 
scattering
data. Its value is not, however, strongly constrained at present
[5]. 
Our aim in this paper is to investigate the influence of this
coupling parameter on the EOS of dense matter and on properties
of neutron stars. To do so
we
adopt here a range of the $\delta$-meson coupling compatible with the Bonn
potentials [5], 
$C_{\delta}^2 \le 4.4fm^2$. The maximum value of $C_{\delta}^2$ we use
exceeds  the  value corresponding to the Bonn potential C [5]
which is $C_{\delta}^2 \approx 2.5 fm^2$.

For a given $\delta$-coupling, $C_{\delta}^2$, we obtain the $\rho$-meson
coupling, $C_{\rho}^2$, from Eq.(18) applying for the soft EOS
the coefficients $A=0.63$ and $B=5.0fm^2$. For the stiff EOS,
the coefficients in Eq.(18) are $A=0.60$ and $B=4.31fm^2$.

To obtain the EOS of the neutron star matter we first calculate the
proton fraction $x$ of the charge-neutral $\beta$-stable neutron star matter, which
satisfies the condition

$$\mu_N-\mu_P = \sqrt{k_N^2+m_N^2}-\sqrt{k_P^2+m_P^2}+{1 \over
2}C_{\rho}^2(1-2x)n_B=\mu_e=\mu_{\mu}, \eq(19) $$ 

\ni where $\mu_e=k_e=(3\pi^2n_e)^{1/3}$ is the electron chemical
potential and $\mu_{\mu}=\sqrt{k_{\mu}^2+m_{\mu}^2}$ is the muon
chemical potential. Charge neutrality requires that $n_e+n_{\mu}=n_P=xn_B$.

In Fig.2 we show the proton fraction $x$ as a function of
density for a few
values of $C_{\delta}^2$.  
One can notice that for both EOS's the proton fraction is
substantially larger for indicated values of the $\delta$-meson coupling,
$C_{\delta}^2$,  than for vanishing coupling, $C_{\delta}^2=0$.
It exceeds the critical value for the direct URCA process to dominate
the cooling rate of neutron stars, which is 
$x_{URCA}\approx 0.11$, already at densities less than twice the
nuclear saturation density. 
 For the stiff EOS and for
the $\delta$-coupling 
corresponding to the Bonn potential C, $C_{\delta}^2=2.5fm^2$,
the proton fraction is about  $40\%$ higher than for vanishing
$\delta$-meson coupling.
For the soft
EOS the proton fraction increases by about $20\%$, a somewhat weaker effect.

The energy density of the $\beta$-stable neutron star matter,
$\epsilon_{ns}$,  is obtained as a
sum of nucleon and lepton contributions,

$$ \epsilon_{ns}=\epsilon_{nuc}+{1 \over 4\pi^2} k_e^4 + {2
\over (2\pi)^3} \int_0^{k_{\mu}}d^3k\sqrt{k^2+m_{\mu}^2}, \eq(20)$$

\ni where nucleon contribution, $\epsilon_{nuc}$, is given in
Eq.(17). Next terms represent the energy density of the electron
Fermi sea and the muon Fermi
sea. The energy per baryon 
with the contribution of the 
$\delta$-field for both soft and stiff EOS is shown in Fig.3.
For comparison, curves with no $\delta$-field
included are also shown. As one can see in
Fig.3 the
energy per particle increases  with
$\delta$-coupling $C_{\delta}^2$. 
The effect is stronger for the stiff EOS. For
$C_{\delta}^2=2.5fm^2$ the energy per baryon is about $10\%$
higher than for $C_{\delta}^2=0$. For the
soft EOS the energy per baryon increases by $\sim 5\%$.

It is interesting to note that actually the
nucleon contribution to the energy per baryon at a given $n_B$
decreases with increasing  
$C_{\delta}^2$ for
both EOS's. It is the lepton contribution which makes the total
energy per particle higher.  

This behaviour can be easily understood. With
increasing $C_{\delta}^2$ the proton fraction of the neutron star
matter in $\beta$-equilibrium increases making the system less
isospin-asymmetric. This in turn reduces the amplitude of the
$\rho$-meson mean field, ${\bar \rho}_0^{(3)}$, which provides a
repulsive contribution. 
As a result, the nucleon energy per baryon is lowered in spite
of the fact that the $\rho$-meson coupling parameter,
$C_{\rho}^2$, is higher. 
However,  
increase of the proton fraction results in higher density of
electrons and muons which thus contribute more to the total
energy per particle.
The energy per baryon of the $\beta$-stable neutron star matter displays a
similar behaviour as the one of pure neutron matter which
becomes higher with increasing $\delta$-meson coupling [2].

In Fig.4 pressure as a function of mass density,
$P=P(\epsilon/c^2)$, a relation referred to as the EOS of neutron star matter,
is shown. In this plot curves corresponding
to $C_{\delta}^2=4.4fm^2$ and to $C_{\delta}=0$ differ
significantly for both  soft and stiff EOS.
Pressure at a given mass 
density is higher for $C_{\delta}^2=4.4fm^2$ than for
$C_{\delta}^2=0$ which indicates that the EOS becomes somewhat
stiffer when the $\delta$-meson contribution is present.

The soft EOS and the stiff
EOS calculated for 
$C_{\delta}^2=2.5 fm^2$, a value corresponding to the Bonn
potential C, are given in Table Ia and Table Ib, respectively.
As the proton fraction at high densities is rather 
high, especially 
for the stiff EOS, leptons (electrons and muons) make a sizable
contribution to the total mass density. 

To gauge the influence of the $\delta$-meson coupling on the EOS
we have calculated models of neutron stars.
The high density EOS calculated in the previous section was matched
with the low density EOS due to 
Baym, Bethe, and Pethick [8] by  constructing a proper
phase transition. By making this construction we have found
interface between crust matter, described by the EOS of Ref.[8],
and liquid core matter, described by our RMF theory.
Both EOS's in Table Ia
and Ib are given for liquid core matter starting from the core
interface density. 

In Fig.5 density profiles are shown for the canonical neutron
star mass $1.4 M_{\odot}$.
For the soft EOS generally the radii are smaller than
for the stiff EOS. The effect of the $\delta$-meson contribution
is more profound in case of the soft EOS. The radius increases from
$R \approx 10.6 km$, for $C_{\delta}^2=0$, to $R \approx 11.9 km$, for
$C_{\delta}^2=2.5 fm^2$, and to $R \approx 12.5 km$, for
$C_{\delta}^2=4.4 fm^2$. For the stiff EOS the radius  is $R
\approx 13.8 km$ for $C_{\delta}^2=0$, $R\approx 14.2 km$ for
$C_{\delta}^2=2.5 fm^2$, and $R \approx 14.6 km$, for 
$C_{\delta}^2=4.4 fm^2$.
 The central density decreases with
increasing $\delta$-coupling. In case of the soft EOS, the
central density is $n_c\approx 8.5n_0$, $n_c \approx 5.4n_0$ and
$n_c \approx 4.4n_0$ for, respectively, $C_{\delta}^2=0$,
$C_{\delta}^2=2.5fm^2$ and $C_{\delta}^2=4.4fm^2$. For the stiff
EOS corresponding central densities are $n_c \approx 2.3n_0$,
$n_c \approx 2.1n_0$ and $n_c \approx 1.9n_0$, respectively.

It is interesting to study the core of $1.4 M_{\odot}$ neutron
star models with the proton fraction exceeding the critical URCA
value, $x_{URCA}$, where the
direct URCA process dominates the cooling rate. With increasing
$C_{\delta}^2$ there are two 
effects influencing its size which tend to cancel one another: 
increase of the proton fraction at a given baryon density
(Fig.2) and decrease of the central density $n_c$ due to
stiffening of the EOS. The former
one tends to extend the core while the latter one tends to
shrink it. As a result, the core size is rather insensitive to changes of
the $\delta$-coupling $C_{\delta}^2$. It is determined primarily
by
the stiffness of the EOS for $C_{\delta}^2=0$. Generally, the mass of the
core where direct URCA process dominates is
higher for the soft EOS than for the stiff EOS.

The neutron star mass as a function of central density is
displayed in Fig.6 for the soft and the stiff EOS. We also show
results for the intermediate EOS, with parameters given in Sect.2. Maximum  
mass
increases slightly with $C_{\delta}^2$. 
The $\delta$-field plays a more important role for the soft
EOS. The maximum  neutron star mass is $M_{max}=1.403
M_{\odot}$ for $C_{\delta}^2=0$. Since this value is less than
the mass of the neutron star in the binary pulsar PSR B1913+16,
which is $1.44 M_{\odot}$, this EOS is too soft to be realistic.
However, with inclusion of the $\delta$-field contribution the
maximum neutron star mass increases to $M_{max}=1.452
M_{\odot}$, for $C_{\delta}^2=2.5fm^2$, and to $M_{max}=1.48
M_{\odot}$, for $C_{\delta}^2=4.4 fm^2$. Inclusion of the
$\delta$-meson with $C_{\delta}^2=2.5 fm^2$ results in about
$4\%$ increase of $M_{max}$ that 
makes the soft EOS astrophysically acceptable. 

For the stiff EOS the maximum  neutron star mass is
$M_{max}=2.275 M_{\odot}$, $M_{max}=2.309 M_{\odot}$ and $M_{max}=2.313
M_{\odot}$ for, respectively, $C_{\delta}^2=0$, $C_{\delta}^2=2.5
fm^2$ and $C_{\delta}^2=4.4 fm^2$. In this case $M_{max}$
increases by $\sim 1.5\%$ for $C_{\delta}^2=2.5fm^2$. Central
densities of the above maximum mass neutron stars are,
respectively, $n_c \approx 6.2n_0$, $n_c \approx 5.9n_0$ and
$n_c \approx 5.7n_0$ 

\bs

\ni{\bf 3. Conclusions and discussion}

We have studied the influence of the $\delta$-meson coupling on
the EOS of neutron star matter in the RMF theory. When the
isovector scalar field of the $\delta$-meson is added to the
standard RMF model, the nuclear symmetry energy has two
contributions of opposite sign. The conventional $\rho$-meson
contribution is positive, whereas the $\delta$-meson
contribution is negative [2]. This reflects the fact that in
pure neutron matter the $\rho$-meson provides repulsion whereas
the $\delta$-meson produces additional attraction. It is thus
obvious that in order to make physically relevant predictions of
the EOS of neutron star matter the coupling parameters of the
$\delta$ and $\rho$ mesons to nucleons should be constrained in such a way
that the empirical value of the nuclear symmetry is preserved.

We have found that the quantity most strongly affected by the
presence of the $\delta$-field is the proton fraction of the
neutron star matter in $\beta$-equilibrium. It increases rapidly
with $C_{\delta}^2$. The effect is more
profound for the stiff EOS, for which the proton fraction
increases by $\sim 40\%$ for the value of $C_{\delta}^2=2.5
fm^2$ corresponding to the Bonn potential C. For the soft
EOS the proton fraction increases in this case by $\sim 20\%$.
For higher values of the $\delta$-coupling parameter this
increase is larger (Fig.2). The URCA
threshold concentration of protons, $x_{URCA} \approx 0.11$, occurs at
lower densities.

The $\delta$-field contribution makes the energy per baryon of
$\beta$-stable neutron star matter
higher by $\sim 10\%$ and $\sim 5\%$ for the stiff and soft EOS,
respectively, and for the Bonn potential C value of
$C_{\delta}^2$. Here the increase is due to  larger leptonic
contribution since the nucleon contribution actually decreases
with $C_{\delta}^2$. The EOS with $C_{\delta}^2\ne 0$ is
slightly stiffer than for $C_{\delta}^2=0$. Maximum mass of the
neutron star also increases by a few percent.

One should stress that the extension of the RMF model to include the
isovector scalar meson $\delta$ is the most natural one. In
fact, the standard RMF model with no $\delta$-field is not fully
selfconsistent, as for isospin-asymmetric matter the density
$<{\bar \psi}\tau_3\psi> \ne 0 $ 
whereas the mean field with the same quantum numbers vanishes.
The coupling constant $g_{\delta}$ is not well determined at
present [5]. However for values in the range compatible with the
Bonn potentials [5] the contribution of the $\delta$-field in
pure neutron matter at saturation density is quite strong [2].
If the $\rho$-coupling is unchanged, the neutron matter becomes
selfbound for  $C_{\delta}^2\approx
1.0fm^2$. To avoid such an unphysical behaviour the
$\rho$-coupling  $C_{\rho}^2$ should be readjusted to meet the
requirement that the
empirical value of the nuclear symmetry energy is reproduced.

\bs
\bs

\ni {\bf Acknowledgements}

This research is partially supported by the Polish State
Committee for Scientific Research (KBN), grants 
2 P03D 001 09 and 2 P03B 083 08.

\bs

\ni {\bf References}

\ni~~[1]~B. D. Serot and J. D. Walecka, Adv.  Nucl.
Phys. {\bf16} (1986) 1.

\ni~~[2]~S. Kubis and M. Kutschera, 
Phys.Lett.B {\bf 399} (1997) 191.

\ni~~[3]~J. Boguta and A. Bodmer, Nucl. Phys. {\bf A292} (1977) 413.

\ni~~[4]~W. D. Myers and W. D. Swiatecki, Ann. Phys. {\bf 84}
(1973) 186; 

\ni~~~~~~J. M. Pearson, Y. Aboussir, A. K. Dutta, R. C. Nayak,
M. Farine, and F. Tondeur, 

\ni~~~~~~Nucl. Phys. {\bf A528} (1991) 1;

\ni~~~~~~P. M\"oller and J. R. Nix, At. Data and Nucl. Data
Tables {\bf 39} (1988) 219;

\ni~~~~~~P. M\"oller, W. D. Myers, W. J. Swiatecki, and J.
Treiner, At. Data and Nucl. Data 

\ni~~~~~~Tables {\bf 39} (1988) 225;

\ni~~~~~~W. D. Myers, W. J. Swiatecki, T. Kodama, L. J.
El-Jaick, and E. R. Hilf, 

\ni~~~~~Phys. Rev. {\bf C15} (1977) 2032.

\ni~~[5]~R. Machleidt, Adv. Nucl. Phys. {\bf 19} (1989) 189.

\ni~~[6]~M. Kutschera, astro-ph/9612143, in "Solar Astrophysics,
Structure of Neutron Stars, 

\ni~~~~~~Gamma Flashes. Meeting of the
Commission for Astrophysics of the Polish
Academy 

\ni~~~~~~of Arts and Sciences", ed. K. Grotowski, Cracow, 1997.

\ni~~[7]~~J. H. Taylor and J. M. Weisberg, Astrophys. J. {\bf
345} (1989) 434.

\ni~~[8]~G. Baym, H. A. Bethe and C. J. Pethick, Nucl. Phys.
{\bf A175} (1971) 225.
\bs
\vfill
\break

\ni $~~~~~~~~~~~~~~~~~~~~~~~~~~~~~~ $ Table Ia

\ni $~~~~~~~~~~~~ $ Soft equation of state, $C_{\delta}^2=2.5fm^2$

\ni ==============================

\ni $~~~~~~~n_B(cm^{-3})~~~~~~~~~ \rho(g/cm^3)~~~~~~~~ P(dynes/cm^2)$

\ni --------------------------------------------------------------------

 8.320$\times 10^{37}$~~~~~~~1.404$\times 10^{14}$~~~~~~~1.612$\times 10^{31}$

 8.360$\times 10^{37}$~~~~~~~1.411$\times 10^{14}$~~~~~~~4.034$\times 10^{31}$

 8.400$\times 10^{37}$~~~~~~~1.418$\times 10^{14}$~~~~~~~6.487$\times 10^{31}$

 8.440$\times 10^{37}$~~~~~~~1.425$\times 10^{14}$~~~~~~~8.971$\times 10^{31}$

 8.481$\times 10^{37}$~~~~~~~1.431$\times 10^{14}$~~~~~~~1.148$\times 10^{32}$

 8.562$\times 10^{37}$~~~~~~~1.445$\times 10^{14}$~~~~~~~1.661$\times 10^{32}$

 8.603$\times 10^{37}$~~~~~~~1.452$\times 10^{14}$~~~~~~~1.922$\times 10^{32}$

 8.644$\times 10^{37}$~~~~~~~1.459$\times 10^{14}$~~~~~~~2.186$\times 10^{32}$

 8.686$\times 10^{37}$~~~~~~~1.466$\times 10^{14}$~~~~~~~2.454$\times 10^{32}$

 9.248$\times 10^{37}$~~~~~~~1.561$\times 10^{14}$~~~~~~~6.288$\times 10^{32}$

 1.013$\times 10^{38}$~~~~~~~1.712$\times 10^{14}$~~~~~~~1.303$\times 10^{33}$

 1.077$\times 10^{38}$~~~~~~~1.821$\times 10^{14}$~~~~~~~1.840$\times 10^{33}$

 1.217$\times 10^{38}$~~~~~~~2.061$\times 10^{14}$~~~~~~~3.161$\times 10^{33}$

 1.334$\times 10^{38}$~~~~~~~2.263$\times 10^{14}$~~~~~~~4.407$\times 10^{33}$

 1.462$\times 10^{38}$~~~~~~~2.486$\times 10^{14}$~~~~~~~5.877$\times 10^{33}$

 1.756$\times 10^{38}$~~~~~~~3.003$\times 10^{14}$~~~~~~~9.672$\times 10^{33}$

 1.924$\times 10^{38}$~~~~~~~3.303$\times 10^{14}$~~~~~~~1.211$\times 10^{34}$

 2.109$\times 10^{38}$~~~~~~~3.634$\times 10^{14}$~~~~~~~1.499$\times 10^{34}$

 2.321$\times 10^{38}$~~~~~~~4.019$\times 10^{14}$~~~~~~~1.858$\times 10^{34}$

 2.561$\times 10^{38}$~~~~~~~4.457$\times 10^{14}$~~~~~~~2.295$\times 10^{34}$

 2.825$\times 10^{38}$~~~~~~~4.945$\times 10^{14}$~~~~~~~2.816$\times 10^{34}$

 3.116$\times 10^{38}$~~~~~~~5.491$\times 10^{14}$~~~~~~~3.436$\times 10^{34}$

 3.437$\times 10^{38}$~~~~~~~6.100$\times 10^{14}$~~~~~~~4.175$\times 10^{34}$

 3.917$\times 10^{38}$~~~~~~~7.026$\times 10^{14}$~~~~~~~5.380$\times 10^{34}$

 4.321$\times 10^{38}$~~~~~~~7.818$\times 10^{14}$~~~~~~~6.485$\times 10^{34}$

 4.766$\times 10^{38}$~~~~~~~8.705$\times 10^{14}$~~~~~~~7.797$\times 10^{34}$

 5.258$\times 10^{38}$~~~~~~~9.700$\times 10^{14}$~~~~~~~9.355$\times 10^{34}$

 5.800$\times 10^{38}$~~~~~~~1.081$\times 10^{15}$~~~~~~~1.120$\times 10^{35}$

 6.619$\times 10^{38}$~~~~~~~1.254$\times 10^{15}$~~~~~~~1.426$\times 10^{35}$

 7.555$\times 10^{38}$~~~~~~~1.456$\times 10^{15}$~~~~~~~1.810$\times 10^{35}$

 8.250$\times 10^{38}$~~~~~~~1.611$\times 10^{15}$~~~~~~~2.121$\times 10^{35}$

 9.841$\times 10^{38}$~~~~~~~1.974$\times 10^{15}$~~~~~~~2.910$\times 10^{35}$

 1.123$\times 10^{39}$~~~~~~~2.305$\times 10^{15}$~~~~~~~3.686$\times 10^{35}$

 1.281$\times 10^{39}$~~~~~~~2.696$\times 10^{15}$~~~~~~~4.670$\times 10^{35}$

 1.463$\times 10^{39}$~~~~~~~3.160$\times 10^{15}$~~~~~~~5.917$\times 10^{35}$

 1.669$\times 10^{39}$~~~~~~~3.711$\times 10^{15}$~~~~~~~7.501$\times 10^{35}$

 1.905$\times 10^{39}$~~~~~~~4.368$\times 10^{15}$~~~~~~~9.515$\times 10^{35}$

 2.175$\times 10^{39}$~~~~~~~5.154$\times 10^{15}$~~~~~~~1.208$\times 10^{36}$

\ni --------------------------------------------------------------------

\vfill
\break

\ni $~~~~~~~~~~~~~~~~~~~~~~~~~~~~~ $ Table Ib

\ni $~~~~~~~~~~~~~ $ Stiff equation of state, $C_{\delta}^2=2.5fm^2$

\ni ==============================

\ni $~~~~~~~n_B(cm^{-3})~~~~~~~~~ \rho(g/cm^3)~~~~~~~~ P(dynes/cm^2)$

\ni --------------------------------------------------------------------

 3.190$\times 10^{37}$~~~~~~~5.359$\times 10^{13}$~~~~~~~8.324$\times 10^{29}$

 3.526$\times 10^{37}$~~~~~~~5.924$\times 10^{13}$~~~~~~~9.026$\times 10^{30}$

 3.898$\times 10^{37}$~~~~~~~6.550$\times 10^{13}$~~~~~~~2.364$\times 10^{31}$

 4.310$\times 10^{37}$~~~~~~~7.242$\times 10^{13}$~~~~~~~4.792$\times 10^{31}$

 4.765$\times 10^{37}$~~~~~~~8.007$\times 10^{13}$~~~~~~~8.641$\times 10^{31}$

 5.268$\times 10^{37}$~~~~~~~8.853$\times 10^{13}$~~~~~~~1.453$\times 10^{32}$

 6.438$\times 10^{37}$~~~~~~~1.082$\times 10^{14}$~~~~~~~3.625$\times 10^{32}$

 7.118$\times 10^{37}$~~~~~~~1.197$\times 10^{14}$~~~~~~~5.486$\times 10^{32}$

 7.869$\times 10^{37}$~~~~~~~1.324$\times 10^{14}$~~~~~~~8.135$\times 10^{32}$

 8.700$\times 10^{37}$~~~~~~~1.465$\times 10^{14}$~~~~~~~1.186$\times 10^{33}$

 1.044$\times 10^{38}$~~~~~~~1.764$\times 10^{14}$~~~~~~~2.282$\times 10^{33}$

 1.255$\times 10^{38}$~~~~~~~2.125$\times 10^{14}$~~~~~~~4.237$\times 10^{33}$

 1.375$\times 10^{38}$~~~~~~~2.335$\times 10^{14}$~~~~~~~5.702$\times 10^{33}$

 1.507$\times 10^{38}$~~~~~~~2.566$\times 10^{14}$~~~~~~~7.555$\times 10^{33}$

 1.652$\times 10^{38}$~~~~~~~2.821$\times 10^{14}$~~~~~~~9.925$\times 10^{33}$

 1.810$\times 10^{38}$~~~~~~~3.104$\times 10^{14}$~~~~~~~1.296$\times 10^{34}$

 1.984$\times 10^{38}$~~~~~~~3.418$\times 10^{14}$~~~~~~~1.683$\times 10^{34}$

 2.175$\times 10^{38}$~~~~~~~3.766$\times 10^{14}$~~~~~~~2.178$\times 10^{34}$

 2.399$\times 10^{38}$~~~~~~~4.183$\times 10^{14}$~~~~~~~2.855$\times 10^{34}$

 2.646$\times 10^{38}$~~~~~~~4.652$\times 10^{14}$~~~~~~~3.727$\times 10^{34}$

 2.919$\times 10^{38}$~~~~~~~5.180$\times 10^{14}$~~~~~~~4.846$\times 10^{34}$

 3.219$\times 10^{38}$~~~~~~~5.777$\times 10^{14}$~~~~~~~6.277$\times 10^{34}$

 3.551$\times 10^{38}$~~~~~~~6.454$\times 10^{14}$~~~~~~~8.102$\times 10^{34}$

 3.917$\times 10^{38}$~~~~~~~7.225$\times 10^{14}$~~~~~~~1.042$\times 10^{35}$

 4.321$\times 10^{38}$~~~~~~~8.105$\times 10^{14}$~~~~~~~1.338$\times 10^{35}$

 4.766$\times 10^{38}$~~~~~~~9.114$\times 10^{14}$~~~~~~~1.713$\times 10^{35}$

 5.258$\times 10^{38}$~~~~~~~1.027$\times 10^{15}$~~~~~~~2.188$\times 10^{35}$

 5.800$\times 10^{38}$~~~~~~~1.161$\times 10^{15}$~~~~~~~2.788$\times 10^{35}$

 6.619$\times 10^{38}$~~~~~~~1.378$\times 10^{15}$~~~~~~~3.852$\times 10^{35}$

 7.555$\times 10^{38}$~~~~~~~1.643$\times 10^{15}$~~~~~~~5.276$\times 10^{35}$

 8.622$\times 10^{38}$~~~~~~~1.971$\times 10^{15}$~~~~~~~7.183$\times 10^{35}$

 9.841$\times 10^{38}$~~~~~~~2.381$\times 10^{15}$~~~~~~~9.716$\times 10^{35}$

 1.123$\times 10^{39}$~~~~~~~2.895$\times 10^{15}$~~~~~~~1.304$\times 10^{36}$

 1.281$\times 10^{39}$~~~~~~~3.542$\times 10^{15}$~~~~~~~1.740$\times 10^{36}$

 1.463$\times 10^{39}$~~~~~~~4.357$\times 10^{15}$~~~~~~~2.307$\times 10^{36}$

 1.669$\times 10^{39}$~~~~~~~5.390$\times 10^{15}$~~~~~~~3.044$\times 10^{36}$

 1.905$\times 10^{39}$~~~~~~~6.700$\times 10^{15}$~~~~~~~4.000$\times 10^{36}$

 2.175$\times 10^{39}$~~~~~~~8.366$\times 10^{15}$~~~~~~~5.239$\times 10^{36}$

\ni -------------------------------------------------------------------

\vfill
\break
\epsffile{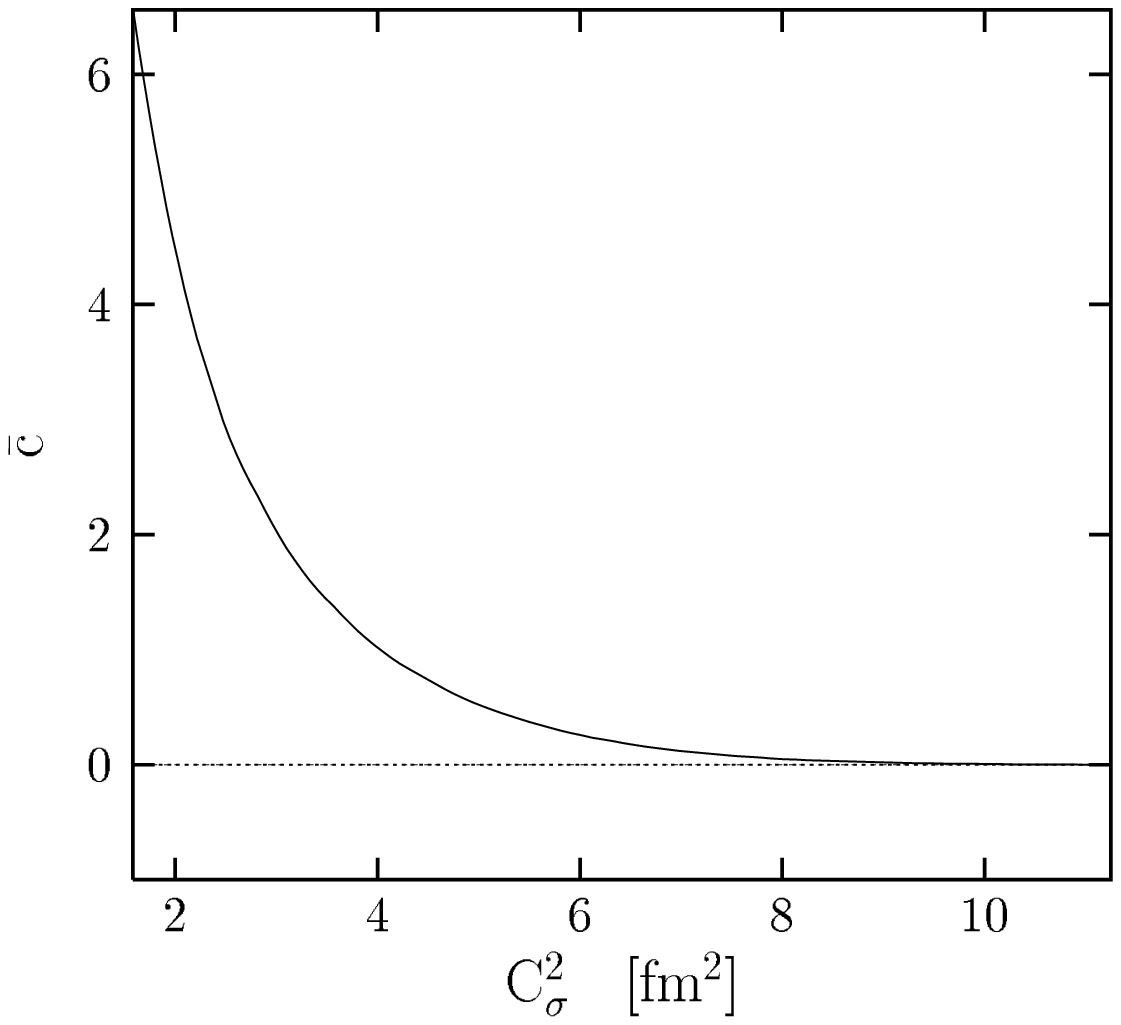}

\ni Fig.1

\ni The coupling parameter, ${\bar c}$, as a function of the
coupling parameter $C_{\sigma}^2$.

\vfill
\break

\epsffile{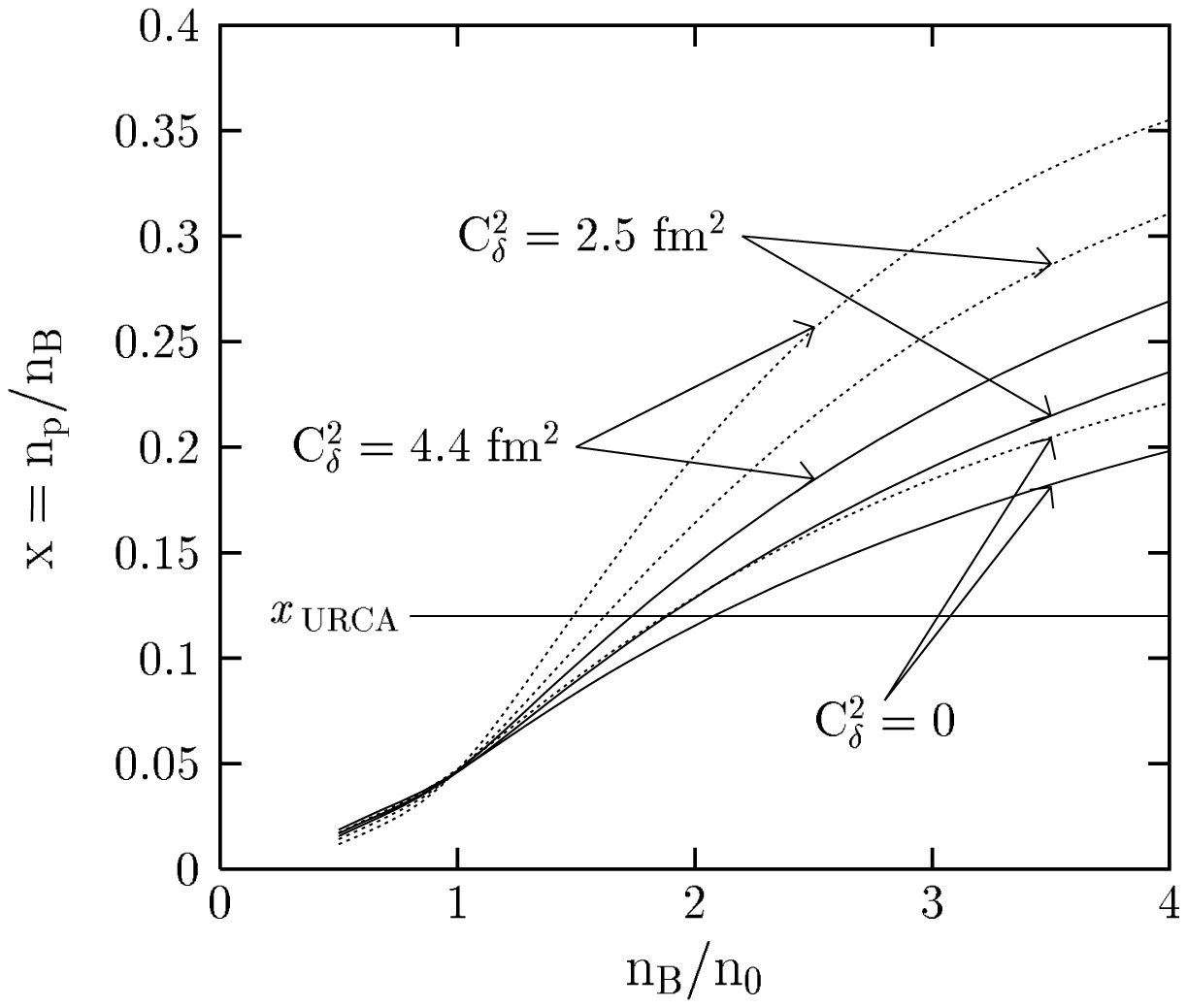}
\ni Fig.2

\ni The proton fraction, $x$, of the neutron star matter for the
soft EOS (solid
curves) and the stiff EOS (dashed curves), for indicated values of
the $\delta$-meson coupling, $C_{\delta}^2$.
\vfill
\break

\epsffile{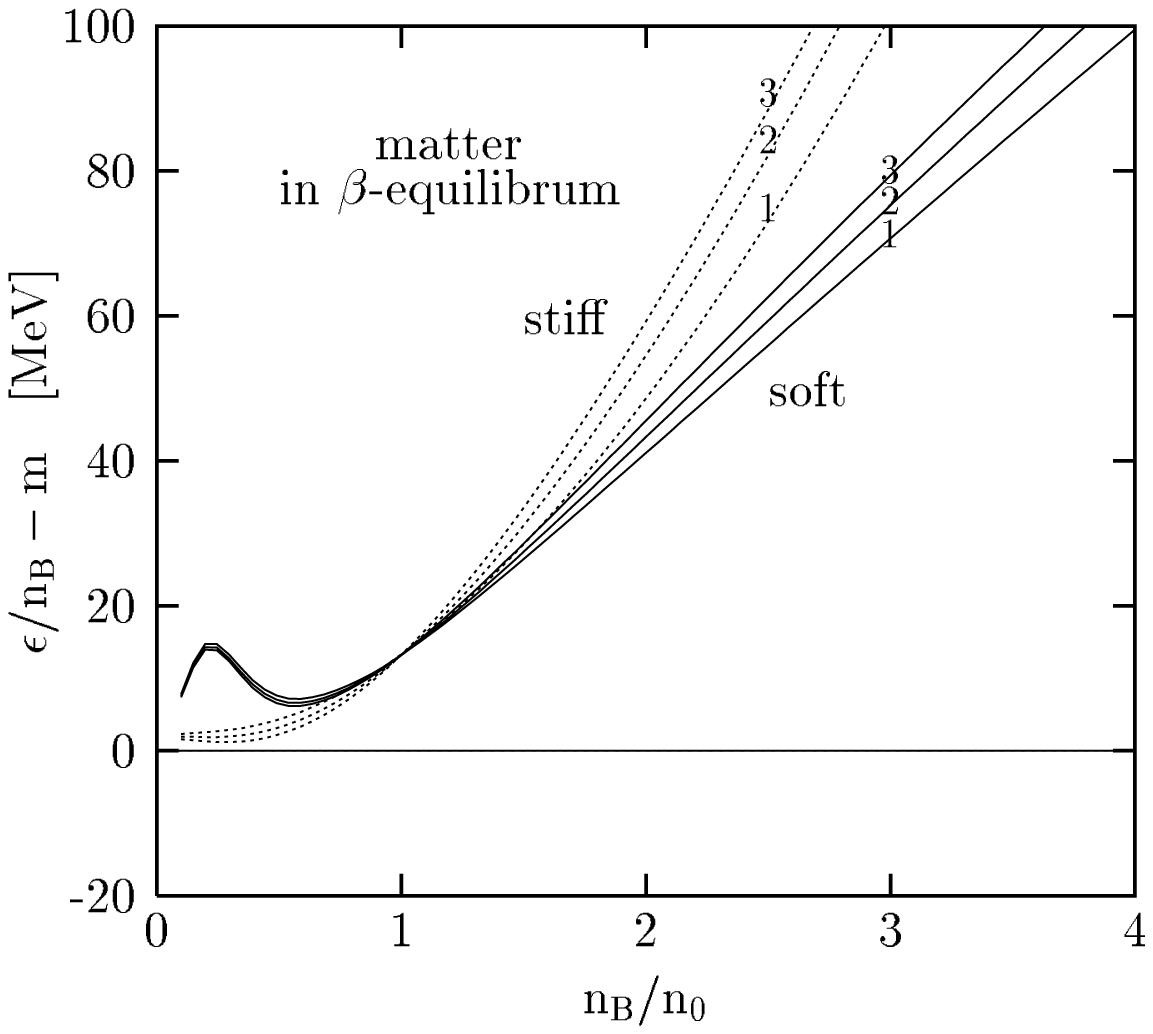}
\ni Fig.3

\ni The energy per particle of the neutron star matter for the
soft EOS and the
stiff EOS. Curves labeled 1, 2, and 3 correspond, respectively, to
$C_{\delta}^2=0$, $C_{\delta}^2=2.5fm^2$, and $C_{\delta}^2=4.4fm^2$.
\vfill
\break

~~~~~~~~

\bs
\bs\bs\bs\bs\bs\bs

~~~

\epsffile{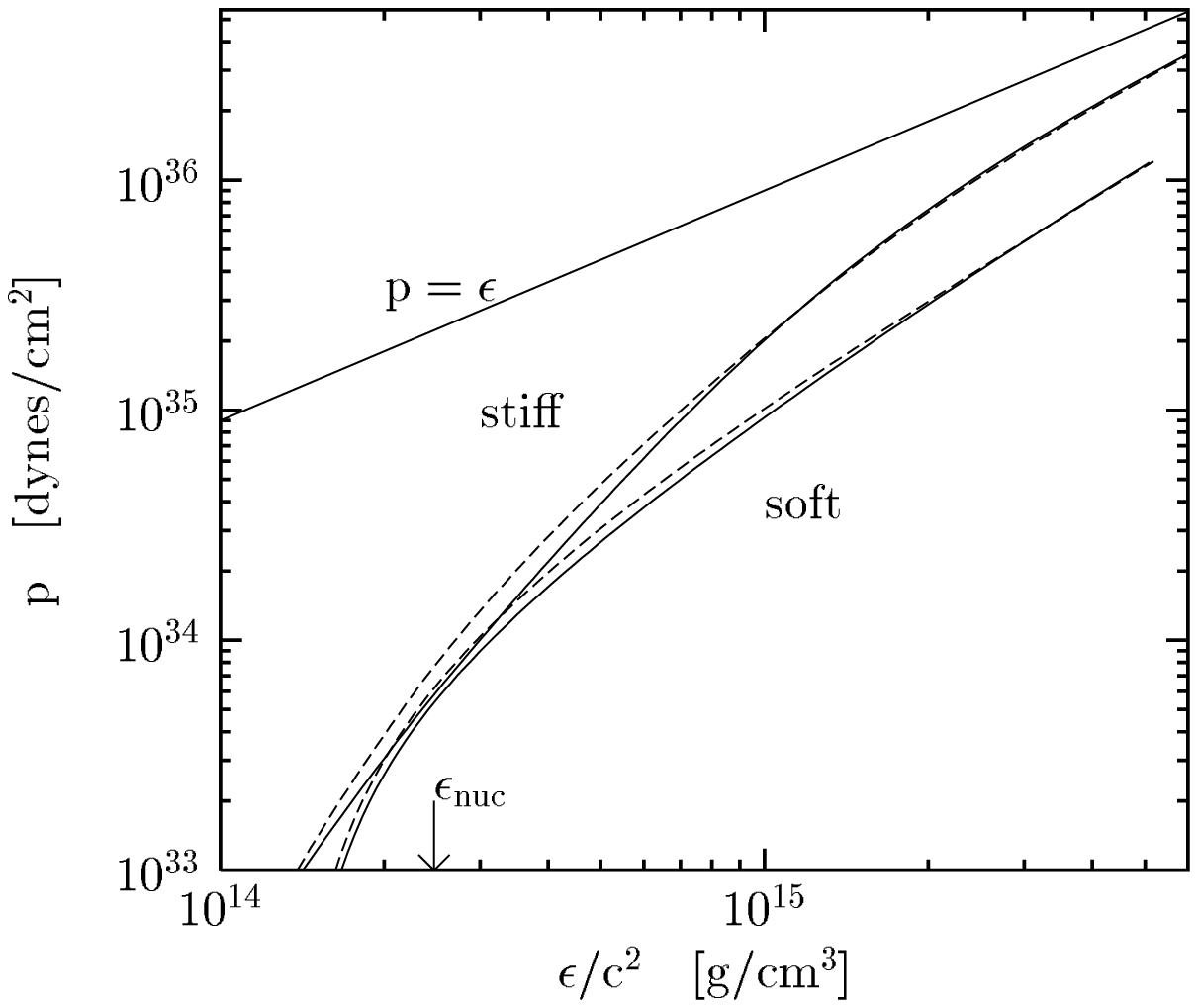}
\ni Fig.4

\ni Pressure as a function of mass density for the soft EOS and
the stiff
EOS. Solid and dashed lines correspond, respectively, to
$C_{\delta}^2=0$ and $C_{\delta}^2=4.4fm^2$.
\vfill
\break

\epsffile{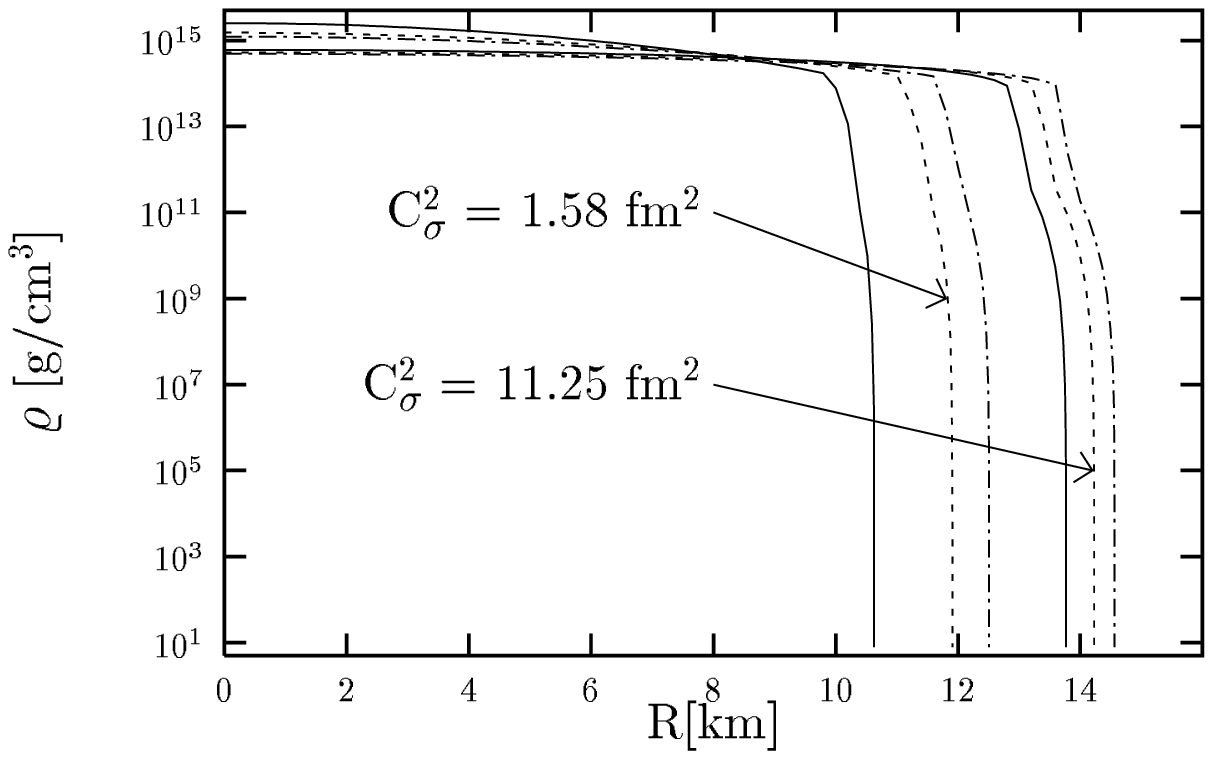}
\ni Fig.5

\ni The density profile of $1.4 M_{\odot}$ neutron star for the
soft EOS and for the 
stiff EOS. Solid, dashed and dotted-dashed lines correspond,
respectively, to $C_{\delta}^2=0$, $C_{\delta}^2=2.5 fm^2$, and
$C_{\delta}^2=4.4fm^2$. 
\vfill
\break

\epsffile{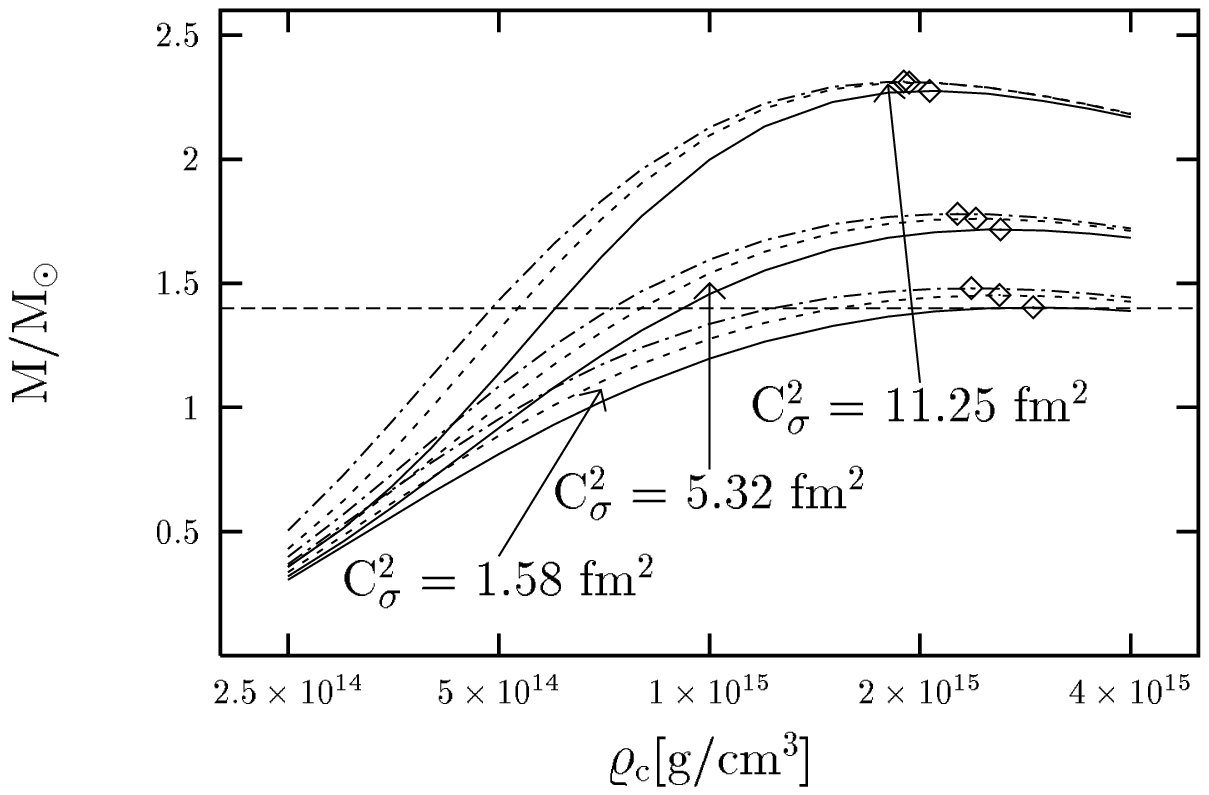}
\ni Fig.6

\ni The spectrum of neutron star masses for the soft EOS and the
stiff EOS. Also, results for the intermediate EOS are shown.
Solid, dashed and dotted-dashed lines correspond, 
respectively, to $C_{\delta}^2=0$, $C_{\delta}^2=2.5 fm^2$, and
$C_{\delta}^2=4.4fm^2$. 

\vfill
\end